\documentclass[manuscript]{aastex}

\usepackage{url}
\usepackage{lscape}
\usepackage{latexsym}
\usepackage{graphics}
\usepackage{graphicx}
\usepackage{enumitem}
\usepackage{amsmath}
\usepackage{mathptmx}
\usepackage{latexsym}
\usepackage[colorlinks,citecolor=blue,urlcolor=blue,linkcolor=blue]{hyperref}
\usepackage{fix-cm}
\usepackage{bigstrut}
\usepackage{booktabs}
\usepackage{array}
\usepackage{breqn}

\shorttitle{A new class of solutions}
\shortauthors{Thomas and Pandya \textit{et al.}}

\begin{document}

\title{A new class of solutions of compact stars with charged distributions on pseudo-spheroidal spacetime}

\author{V. O. Thomas\altaffilmark{1}}
\affil{Department of Mathematics, Faculty of Science, The Maharaja Sayajirao University of Baroda,\\ Vadodara - 390 002, India}
\email{votmsu@gmail.com}

\and

\author{D. M. Pandya\altaffilmark{2}}
\affil{Department of Mathematics \& Computer Science, Pandit Deendayal Petroleum University, Raisan, Gandhinagar - 382 007, India}
\email{dishantpandya777@gmail.com}


\begin{abstract}

In this paper a new class of exact solutions of Einstein's field equations for compact stars with charged distributions is obtained
on the basis of pseudo-spheroidal spacetime characterized by the metric potential $g_{rr}=\frac{1+K\frac{r^{2}}{R^{2}}}{1+\frac{r^{2}}{R^{2}}}$,
where $K$ and $R$ are geometric parameters of the spacetime. The expressions for radial pressure ($ p_r $) and electric field intensity ($ E $) are chosen in such a way that the model falls in the category of physically acceptable one. The bounds of geometric parameter $K$ and the physical parameters $ p_0 $ and $\alpha$ are obtained by imposing the physical requirements and regularity conditions. The present model is in good agreement with the observational data of various compact stars like 4U 1820-30, PSR J1903+327, 4U 1608-52, Vela X-1, SMC X-4, Cen X-3 given by Gangopadhyay {\em{et al.}} [Gangopadhyay T., Ray S., Li X-D., Dey J. and Dey M., {\it Mon. Not. R. Astron. Soc.} {\bf431} (2013) 3216]. When $ \alpha = 0, $ the model reduces to the uncharged anisotropic distribution given by Thomas and Pandya [Thomas V. O. and Pandya D. M.,~arXiv:1506.08698v1 [gr-qc](26 Jun 2015)]. 

\end{abstract}


\section{Introduction}

The equilibrium of a spherical distribution of matter in the form of perfect fluid is maintained by the repulsive pressure force against the gravitational attraction. For matter distribution in the form of dust, there is no such force to counter the gravitational attraction. In such situations, the collapse of the distribution to a singularity can be averted if the matter is accompanied by some electric charge. The Coulombian force of repulsion due to the presence of charge contributes additional force to the fluid pressure when the matter is in the form of perfect fluid.

A systematic study of electromagnetic fields in the context of general relativity was due to \cite{Rainich25}. The equilibrium of charged dust spheres within the frame work of general relativity was examined critically by \cite{Papapetrou47} and \cite{Majumdar47}. \cite{Bonner60, Bonner65} has shown that a spherical distribution of matter can keep its equilibrium if it is accompanied by certain modest electric charge. \cite{Stettner73} has shown that a uniform density fluid distribution accompanied by some surface charge is more stable than the one without charge. \cite{Krori75} obtained a singularity free solution for static charged fluid spheres. This solution has been analysed in detail by \cite{Juvenicus76}. The solution obtained by \cite{Pant79} for static spherically symmetric relativistic charged  fluid sphere has Tolman Solution VI as a particular case in the absence of charge.

\cite{Cooperstock78} have studied relativistic spherical distributions of charged perfect fluids in equilibrium and obtained explicit solutions of Einstein - Maxwell equations in the interior of a sphere containing uniformly charged dust in equilibrium. \cite{Bonnor75} have obtained a static interior dust metric with matter density increasing outward. \cite{Whitman81} have given a method for solving coupled Einstein - Maxwell equations for spherically symmetric static systems containing charge, obtained a number of analytic solutions and examined their stability. Conformally flat interior solutions were obtained by \cite{Chang83} for charged fluid as well as charged dust distributions.

\cite{Tikekar84} has studied some general aspects of spherically symmetric static distributions of charged fluids for specific choice of density and pressure. This solution admits \cite{Pant79} solution as a particular case. \cite{Patel95} have obtained solutions of Einstein - Maxwell equations. \cite{Rao00} have developed a formalism for generating new solutions of coupled Einstein - Maxwell equations.

The study of charged superdense star models compatible with observational data has generated deep interest among researchers in the recent past and a number of articles have been appeared in this direction \cite{Maurya11a, Maurya11b, Maurya11c, Pant12,Maurya15}. Theoretical investigations of \cite{Ruderman72} and \cite{Canuto74} suggest that matter may not be isotropic in high density regime and hence it is pertinent to study charged models incorporating anisotropy in pressure. Relativistic models of charged fluids distributions on spacetimes with spheroidal geometry have been studied by \cite{Patel87}, \cite{Singh98}, \cite{Sharma01}, \cite{Gupta05} and \cite{Komathiraj07}. 

Charged strange and quark star models have been studied by \cite{Sharma06}, and \cite{Mukherjee01, Mukherjee02}. The study of charged fluid distributions have been carried out recently by \cite{Maurya11a, Maurya11b, Maurya11c}, \cite{Pant12} and \cite{Maurya15}.

In the present article, we have obtained a new class of solutions for charged fluid distribution on the background of pseudo spheroidal spacetime. Particular choices for radial pressure $ p_r $ and electric field intensity $ E $ are taken so that the physical requirements and regularity conditions are not violated. The bounds for the geometric parameter $ K $ and the parameter $ \alpha $ associated with charge, are determined using various physical requirements that are expected to satisfy in its region of validity. It is found that these models can accommodate a number of pulsars like like 4U 1820-30, PSR J1903+327, 4U 1608-52, Vela X-1, SMC X-4, Cen X-3, given by \cite{Gangopadhyay13}. When $ \alpha = 0, $ the model reduces to the uncharged anisotropic distribution given by \cite{Thomas15}.

In section \ref{sec:2}, we have solved Einstein - Maxwell equations and in section \ref{sec:3} the bounds for model parameters $ K $ and $ \alpha $ are obtained using physical acceptability and regularity conditions. In section \ref{sec:4}, we have displayed a variety of pulsars in agreement with the charged pseudo-spheroidal model developed. In section \ref{sec:5} we have examined various physical conditions throughout the distribution through the aid of numerical and graphical methods.

\section{Spacetime Metric}
\label{sec:2}
A three-pseudo spheroid immersed in four-dimensional Euclidean space has the Cartesian equation

\begin{eqnarray}
 \frac{u^2}{b^2} - \frac{x^2 + y^2 + z^2}{R^2} = 1. \nonumber
 \label{three-pseudo spheroid equation}
\end{eqnarray}

The sections $ u = constant $ are spheres of real or imaginary radius according as $ u^2 > b^2 $ or $ u^2 < b^2, $ while the sections $ x = const,~y = const $, and $ z = const $ are respectively, hyperboloids of two sheets. \\ 
On taking the parametrization 

\begin{eqnarray}\label{parametrization}
 \nonumber x & = & R sinh\lambda sin\theta cos\phi \\ \nonumber
 y & = & R sinh\lambda sin\theta sin\phi \\ \nonumber
 z & = & R sinh\lambda cos\theta \\
 u & = & b cosh\lambda,		
\end{eqnarray}
the Euclidean metric
\begin{equation}
 d\sigma^2 = dx^2 + dy^2 + dz^2 + du^2 \nonumber
 \label{sigma}
\end{equation}
takes the form
\begin{equation}
 d\sigma^2 = \frac{1 + K \frac{r^2}{R^2}}{1 + \frac{r^2}{R^2}} dr^2 + r^2 d\theta^2 + r^2 sin^2 \theta d\phi^2
 \label{sigma in spherical coordinates}
\end{equation}

where $ K = 1 + \frac{b^2}{R^2} $ and $ r = R sinh\lambda $. The metric (\ref{sigma in spherical coordinates}) is regular for all points with $ K > 1 $ and call pseudo-spheroidal metric (\cite{Tikekar98}). \\
We shall take the interior spacetime metric representing charged anisotropic matter distribution as
\begin{equation}\label{IMetric1}
	ds^{2}=e^{\nu(r)}dt^{2}-\left(\frac{1+K\frac{r^{2}}{R^{2}}}{1+\frac{r^2}{R^{2}}} \right)dr^{2}-r^{2}\left(d\theta^{2}+\sin^{2}\theta d\phi^{2} \right),
\end{equation}
where $K$ and $R$ are geometric parameters and $K>1$. This spacetime, known as pseudo-spheroidal spacetime, has been studied by number
of researchers~\cite {Tikekar98,Tikekar99,Tikekar05,Thomas05,Thomas07,Paul11,Chattopadhyay10,Chattopadhyay12} and have found that it can accommodate compact superdense stars.

Since the metric potential $g_{rr}$ is chosen apriori, the other metric potential $\nu\left(r \right)$ is to be determined by solving
the Einstein-Maxwell field equations
\begin{equation}\label{FE}
	R_{i}^{j}-\frac{1}{2}R\delta_{i}^{j}=8\pi\left(T_{i}^{j}+\pi_{i}^{j}+E_{i}^{j} \right),
\end{equation}
where,
\begin{equation}\label{Tij}
	T_{i}^{j}=\left(\rho+p \right)u_{i}u^{j}-p\delta_{i}^{j},
\end{equation}
\begin{equation}\label{piij}
	\pi_{i}^{j}=\sqrt{3}S\left[c_{i}c^{j}-\frac{1}{2}\left(u_{i}u^{j}-\delta_{i}^{j} \right) \right],
\end{equation}
and
\begin{equation}\label{Eij}
	E_{i}^{j}=\frac{1}{4\pi}\left(-F_{ik}F^{jk}+\frac{1}{4}F_{mn}F^{mn}\delta_{i}^{j} \right).
\end{equation}
Here $\rho$, $p$, $u_{i}$, $S$ and $c^{i}$, respectively, denote the proper density, fluid pressure, unit-four velocity, magnitude
of anisotropic tensor and a radial vector given by $\left(0, -e^{-\lambda/2}, 0, 0 \right)$. $F_{ij}$ denotes the anti-symmetric
electromagnetic field strength tensor defined by 
\begin{equation}\label{Fij}
	F_{ij}=\frac{\partial A_{j}}{\partial x_{i}}-\frac{\partial A_{i}}{\partial x_{j}},
\end{equation}
which satisfies the Maxwell equations
\begin{equation}\label{ME1}
	F_{ij,k}+F_{jk,i}+F_{ki,j}=0,
\end{equation}
and
\begin{equation}\label{ME2}
	\frac{\partial}{\partial x^{k}}\left(F^{ik}\sqrt{-g} \right)=4\pi\sqrt{-g}J^{i},
\end{equation}
where $g$ denotes the determinant of $g_{ij}$, $A_{i}=\left(\phi(r), 0, 0, 0 \right)$ is four-potential and
\begin{equation}\label{Ji}
	J^{i}=\sigma u^{i},
\end{equation}
is the four-current vector and $\sigma$ denotes the charge density.

The only non-vanishing components of $F_{ij}$ is $F_{01}=-F_{10}$. Here
\begin{equation}\label{F01}
	F_{01}=-\frac{e^{\frac{\nu+\lambda}{2}}}{r^{2}}\int_{0}^{r} 4\pi r^{2}\sigma e^{\lambda/2}dr,
\end{equation}
and the total charge inside a radius $r$ is given by 
\begin{equation}\label{qr}
	q(r)=4\pi\int_{0}^{r} \sigma r^{2}e^{\lambda/2}dr.
\end{equation}
The electric field intensity $E$ can be obtained from $E^{2}=-F_{01}F^{01}$, which subsequently reduces to
\begin{equation}\label{E}
	E=\frac{q(r)}{r^{2}}.
\end{equation}
The field equations given by (\ref{FE}) are now equivalent to the following set of the non-linear ODE's
\begin{equation}\label{FE1}
	\frac{1-e^{-\lambda}}{r^{2}}+\frac{e^{-\lambda}\lambda'}{r}=8\pi\rho+E^{2},	
\end{equation}
\begin{equation}\label{FE2}
	\frac{e^{-\lambda}-1}{r^{2}}+\frac{e^{-\lambda}\nu'}{r}=8\pi p_{r}-E^{2},
\end{equation}
\begin{equation}\label{FE3}
	e^{-\lambda}\left(\frac{\nu{''}}{2}+\frac{\nu{'}^{2}}{4}-\frac{\nu'\lambda'}{4}+\frac{\nu'-\lambda'}{2r} \right)=8\pi p_{\perp}+E^{2},
\end{equation}
where we have taken
\begin{equation}\label{pr1}
	p_{r}=p+\frac{2S}{\sqrt{3}},
\end{equation}
\begin{equation}\label{pp1}
	p_{\perp}=p-\frac{S}{\sqrt{3}}.
\end{equation}
Because $e^{\lambda}=\frac{1+K\frac{r^{2}}{R^{2}}}{1+\frac{r^{2}}{R^{2}}}$, the metric potential $\lambda$ is a known function of $r$. The set of equations (\ref{FE1}) - (\ref{FE3}) are to
be solved for five unknowns $\nu$, $\rho$, $p_{r}$, $p_{\perp}$ and $E$. So we have two free variables for which suitable assumption
can be made. We shall assume the following expressions for $p_{r}$ and $E$ with  the central pressure $ p_0 > 0. $
\begin{equation}\label{pr2}
	8\pi p_{r}=\frac{p_0 \left(1-\frac{r^4}{R^4}\right)}{R^2 \left(1+K\frac{r^2}{R^2}\right)^2},
\end{equation} 
\begin{equation}\label{E2}
	E^{2}=\frac{\alpha~r^2}{R^4}.
\end{equation}
The expressions for $ p_r $ and $ E^2 $ are so selected that it may comply with the physical requirement. A physically acceptable radial pressure $ p_r $ should be finite at the centre $ r = 0, $ decreasing radially outward and finally vanish at the boundary of the distribution. The gradient of $ p_r $ is given by 

\begin{equation}\label{dprdr}
	8\pi\frac{dp_{r}}{dr}=-\frac{4~p_0~r \left(K+\frac{r^2}{R^2}\right)}{R^4 \left(1+\frac{K r^2}{R^2}\right)^3}.
\end{equation}

It can be noticed from equation (\ref{pr2}) that $ 8\pi p_r |_{r = 0} = \frac{p_0}{R^2}, $ which is a finite quantity at $ r = 0. $ It vanishes at $ r = R, $ which is taken as the boundary radius of the star. Further from equation (\ref{dprdr}) it can be noticed that $ p_r $ is a radially decreasing function of $ r $. For a physically acceptable electric field intensity, $ E(0) = 0 $ and $ \frac{dE}{dr} > 0. $ From equation (\ref{E2}), it is evident that $ E $ is a monotonically increasing function of $ r $.           

On substituting the values of $p_{r}$ and $E^{2}$ in (\ref{FE2}) we obtain, after a lengthy 
calculation
\begin{eqnarray}
 \nonumber 	e^{\nu} & = & C \times \exp \left[-\frac{\left(1+\frac{r^2}{R^2}\right) \left(\frac{K^2 \alpha  r^2}{R^2}+(2 p_0+(2-3 K) K \alpha )\right)}{4
   K}\right] R^{(K-1) (1-\alpha )+\frac{(1+K) p_0}{K^2}} \left(1+\frac{r^2}{R^2}\right)^{\frac{1}{2} (K-1) (1-\alpha)} \\ 
   & & \times \left(1+K\frac{r^2}{R^2}\right)^{\frac{(1+K) p_0}{2 K^2}}
   \label{enu}
\end{eqnarray}
where $C$ is a constant of integration.\\

Hence, with the help of the equation (\ref{enu}), spacetime metric (\ref{IMetric1}) can be written explicitly as

\begin{eqnarray}
 \nonumber ds^2 & = & C \times \exp \left[-\frac{\left(1+\frac{r^2}{R^2}\right) \left(\frac{K^2 \alpha  r^2}{R^2}+(2 p_0+(2-3 K) K \alpha )\right)}{4
   K}\right] R^{(K-1) (1-\alpha )+\frac{(1+K) p_0}{K^2}} \left(1+\frac{r^2}{R^2}\right)^{\frac{1}{2} (K-1) (1-\alpha)} \\ 
   &  & \times \left(1+K\frac{r^2}{R^2}\right)^{\frac{(1+K) p_0}{2 K^2}}-\left(\frac{1+K\frac{r^{2}}{R^{2}}}{1+\frac{r^2}{R^{2}}} \right)dr^{2}-r^{2}\left(d\theta^{2}+\sin^{2}\theta d\phi^{2} \right).
   \label{IMetric2}
\end{eqnarray}

The interior spacetime metric (\ref{IMetric2}) is suitable to represent the charged fluid distribution if it match continuously with Riessner-Nordstr{\"o}m metric
\begin{equation}\label{EMetric2}
	ds^{2}=\left(1-\frac{2m}{r}+\frac{q^{2}}{r^{2}} \right)dt^{2}-\left(1-\frac{2m}{r}+\frac{q^{2}}{r^{2}} \right)^{-1}dr^{2}-r^{2}\left(d\theta^{2}+\sin^{2}\theta d\phi^{2}\right),
\end{equation}
across the boundary $r=R$. The continuity of metric coefficients across $ r = R $ provide the estimates of the constant of integration $ C $ and $ M $ as 
\begin{equation}\label{C}
	C =\frac{2}{K + 1} \left(R \sqrt{2}\right)^{(K-1) (\alpha -1)} e^{\frac{p_0}{K}+\alpha  (1-K)} \left((1+K) R^2\right)^{-\frac{(1+K)
   p_0}{2 K^2}}
\end{equation} and 
\begin{equation}\label{M}
	M=\frac{R (\alpha(K+1) + K-1)}{2 (K+1)}.
\end{equation}
Here $M = m(r = R)$ denotes the mass of the star inside the radius $ R $.
\section{Physical Requirements and Bounds for Parameters}
\label{sec:3}
Now, equation (\ref{FE1}) gives the density of the distribution as 
\begin{equation}\label{rho3}
	8\pi\rho=\frac{(K-1) \left(3+ K \frac{r^2}{R^2}\right)}{R^2 \left(1+ K \frac{r^2}{R^2}\right)^2}-\frac{\alpha~r^2}{R^4}.
\end{equation}
The condition $\rho(r=0)>0$ is clearly satisfied and $\rho(r=R)>0$ gives the following inequality connecting $\alpha$ and $K$.
\begin{equation}\label{In1}
	0\leq \alpha <1-\frac{4}{(K+1)^2}.
\end{equation}

Since $ K > 1, $ the inequality (\ref{In1}) implies

\begin{equation}
 0 \leq \alpha < 1.
 \label{first numeric bound on alpha}
\end{equation}

Differentiating (\ref{rho3}) with respect to $r$, we get
\begin{equation}\label{drhodr}
	8\pi\frac{d\rho}{dr}=-\frac{2~r}{R^4} \left( \frac{K(K-1) \left(5+\frac{K r^2}{R^2}\right)+\alpha}{R^4 \left(1+\frac{K r^2}{R^2}\right)^3}\right).
\end{equation}
It is observed that $\frac{d\rho}{dr}(r=0)=0$ and $\frac{d\rho}{dr}(r=R)<0$. In fact $ \rho $ is a decreasing function of $ r $ throughout the distributions.  \\

The expression for $p_{\perp}$ is
\begin{eqnarray}\label{pp3}
\nonumber 8\pi p_{\perp} & = & \frac{1}{2 R^4}\left(\frac{p_0^2 r^2 \left(1-\frac{r^2}{R^2}\right)^2 \left(1+\frac{r^2}{R^2}\right)}{2 \left(1+\frac{K
   r^2}{R^2}\right)^3}+\frac{(-1+K)^2 r^2 \left(3+\frac{K r^2}{R^2}\right)}{2 \left(1+\frac{r^2}{R^2}\right) \left(1+\frac{K
   r^2}{R^2}\right)^2} \right. \\  
   \nonumber & & \left. +\frac{p_0 \left(-K (K+1)\frac{r^6}{R^6}+\left(K^2-3 K-4\right)\frac{r^4}{R^4}-\frac{2
   r^2}{R^2}+2\right)}{R^{10} \left(1+\frac{K r^2}{R^2}\right)^3} \right. \\
   \nonumber & & \left. - \frac{r^2}{R^6}\frac{\left(-p_0 \frac{r^6}{R^6}+ K (K+5)\frac{r^4}{R^4}+(p_0+8K+4)\frac{r^2}{R^2}+6\right) \alpha }{\left(1+\frac{r^2}{R^2}\right) \left(1+\frac{K r^2}{R^2}\right)} \right. \\
   & & \left. + \frac{1}{2}\frac{r^6}{R^4}\frac{\left(1+\frac{K r^2}{R^2}\right) \alpha ^2}{\left(1+\frac{r^2}{R^2}\right)}\right).
\end{eqnarray}

The condition $p_{\perp}>0$ at the boundary $ r = R $ imposes a restriction on $\alpha$ and $ p_0 $ respectively given by 
\begin{equation}\label{In3a}
	0\leq \alpha <\frac{K^2+13 K+10}{(K+1)^2}-\sqrt{\frac{24 K^3+193 K^2+262 K+97}{(K+1)^4}}
\end{equation}
and 
\begin{equation}\label{In3}
	0 < p_0 \leq \frac{1}{16} \left(\alpha ^2 (K+1)^3-2 \alpha  (K (K+13)+10) (K+1)+(K-1)^2 (K+3)\right).
\end{equation}

The anisotropy can be written in the form as 

\begin{equation}
8 \pi \sqrt{3} S = 8 \pi p_r - 8 \pi p_\perp.
\label{anisotropy}
\end{equation}

The expression for $\frac{dp_{\perp}}{dr}$ is given by

\begin{eqnarray}\label{dppdr}
 \nonumber 8 \pi \frac{dp_{\perp}}{dr} & = & \frac{3}{2}\left(K \alpha ^2-\frac{p_0 \left(K^2-1\right) \left(6 K^2-p_0 (K+1)\right)}{K \left(1+\frac{K
   r^2}{R^2}\right)^4}\right)\frac{r^5}{R^8} \\
   \nonumber & & -\left(\frac{2 K^4 (K-1)+\left\{\left(4-K-7 K^2\right) 2 K^2+\left(2 K^3+K^2+1\right)
   p_0\right\} p_0}{K^2 \left(1+\frac{K r^2}{R^2}\right)^3} +\frac{\alpha(K \alpha  (K-1)-2 p_0)}{K}\right) \\
   \nonumber & & \times \frac{r^3}{R^6}+\left(\frac{K^4 (3 K-5)-4 K^3 p_0 (1+3 K)+p_0^2 \left(K^3-1\right)+2 K \alpha  \left(K^2+p_0 (K+1)\right)}{2 K^3 \left(1+\frac{K r^2}{R^2}\right)^2} \right. \\
   \nonumber & & \left. -\frac{(\alpha -1) (3-K+\alpha  (K-1))}{2 \left(1+\frac{r^2}{R^2}\right)^2}+\frac{p_0^2-K \alpha  \left(2 \left(K^3+5 K^2+p_0 (K+1)\right) K \alpha -K^3 \alpha^2 (K-1)\right)}{2 K^3}\right)\\
   & & \times \frac{r}{R^4}.
\end{eqnarray}

Evidently, the value of $\frac{dp_{\perp}}{dr}=0$ at the origin; and at the boundary $\frac{dp_{\perp}}{dr}(r=R)<0$ gives the following bounds for $\alpha$ with $ p_0 > 0 $
\begin{equation}\label{In4a}
	0 \leq \alpha \leq \frac{K(K-3)+10}{(K+1)^2}.
\end{equation}

Using (\ref{first numeric bound on alpha}) in above inequality (\ref{In4a}), we get

\begin{equation}
 K > 1.8
 \label{first numeric bound on K}
\end{equation}

a lower bound for $ K $. \

In order to examine the strong energy condition, we evaluate the expression $\rho-p_{r}-2p_{\perp}$ at the centre and on the boundary
of the star. It is found that, for a positivity of $ \rho - p_r - 2 p_\perp $ at the centre,
\begin{equation}\label{In5}
	0 < p_0 < K-1,
\end{equation}
and $ \left(\rho-p_{r}-2p_{\perp} \right)(r=R)>0 $ gives the bound on $ K $ and $\alpha$, namely
\begin{equation}\label{In6a}
	1 < K \leq 5,
\end{equation}
\begin{equation}\label{In6}
	0\leq \alpha \leq \frac{K^2+11K+8}{(K+1)^2}+\sqrt{\frac{24 K^3+153 K^2+174 K+49}{(K+1)^4}}.
\end{equation}

The expressions for adiabatic sound speed $\frac{dp_{r}}{d\rho}$ and $\frac{dp_{\perp}}{d\rho}$ in the radial and transverse directions, respectively, are given by 
\begin{equation}\label{dprdrho}
	\frac{dp_{r}}{d\rho}=\frac{2 p_0 \left(K+\frac{r^2}{R^2}\right)}{(K-1) K \left(5+\frac{K
   r^2}{R^2}\right)+\left(1+\frac{K r^2}{R^2}\right)^3 \alpha}~,
\end{equation}
and
\begin{equation}\label{dppdrho}
	\frac{dp_{\perp}}{d\rho}= \frac{\frac{dp_\perp}{dr}}{\frac{d\rho}{dr}}
\end{equation}
where $ \frac{d\rho}{dr} $ and $ \frac{dp_\perp}{dr} $ are given by expressions (\ref{drhodr}) and (\ref{dppdr}). \\
The condition $ 0 \leq \frac{dp_{r}}{d\rho}|_{(r = 0)}\leq 1$ gives the following bounds on $ p_0 $ with $ \alpha \geq 0$ and $ K > 1 $, 
\begin{equation}\label{Inpre7}
	0 < p_0\leq \frac{5 (K-1) K + 2 \alpha}{2 K}.
\end{equation}
Moreover, $ 0 \leq \frac{dp_{r}}{d\rho}|_{(r = R)}\leq 1$ leads to the following inequality
\begin{equation}
0 < p_0\leq \alpha  (K+1)^2+\frac{(K-1) K (K+5)}{2 (K+1)}.
 \label{dppdrhoatbdry}
\end{equation}

Further, $ 0 \leq \frac{dp_{\perp}}{d\rho}|_{(r = 0)}\leq 1 $, give the following bounds for $ K $, $ \alpha $ and $ p_0. $
\begin{equation}\label{Inprepre7}
	1 < K < 3.34441,
\end{equation}
\begin{equation}\label{Inpreprepre7}
	0 \leq \alpha < \frac{1}{4} (K-1)^2,
\end{equation}
and 
\begin{equation}
 2 (3 K+1)-\sqrt{12 \alpha +33 K^2+30 K+1} \leq p_0 \leq 2 (3 K+1)-\sqrt{8 \alpha +13 K^2+50 K+1}.
 \label{Inforp0}
\end{equation}

Moreover at the boundary $(r=R)$, we have the following restrictions on $ K, ~\alpha $ and $ p_0 $.

\begin{equation}\label{Inpreprepreprepre7}
	1 < K < 16.4118,
\end{equation}
\begin{equation}\label{Inprepreprepreprepre7}
	0\leq \alpha <\frac{3 K^3+20 K^2+31 K+10}{(K+1)^2 (7 K+5)}-\sqrt{\frac{16 K^6+48 K^5-99 K^4+948 K^3+2054 K^2+1044 K+85}{(K+1)^4 (7 K+5)^2}}
\end{equation}
and
% \begin{dmath}[style={\scriptsize}]
\begin{dmath}
0 < p_0 \leq \frac{1}{8 \alpha +8 \alpha  K^2-8 K^2+16 \alpha  K+24 K-80} \biggl(-5 \alpha ^2+20 \alpha -7 \alpha ^2 K^4+6 \alpha  K^4+K^4-26 \alpha ^2 K^3+46 \alpha  K^3-12 K^3-36 \alpha ^2 K^2+102 \alpha  K^2-78 K^2-22 \alpha ^2 K+82 \alpha  K+92 K-3\biggr).
 \label{p0otherinequality}
\end{dmath}

The necessary condition for the model to represent a stable relativistic star is that $\Gamma>\frac{4}{3}$ throughout the star. $\Gamma>\frac{4}{3}$ 
at $r=0$ gives a bound on $ p_0 $ with $ K > 1 $ and $ \alpha \geq 0 $,  
\begin{equation}
p_0 >\frac{2 \alpha +K^2-K}{3 K}.
\label{inequalityforp0fromgamma}
\end{equation}

The upper limits of $ \alpha $ in the inequalities (\ref{In1}), (\ref{In3a}), (\ref{In4a}), (\ref{In6}), and (\ref{Inpreprepre7}) for different permissible values of $ K $ are shown in Table~\ref{tab:1}. It can be noticed that the smallest bound for $ \alpha $ is given by (\ref{In3a}). \\ 
The lower bounds of $ p_0 $ are calculated from (\ref{Inforp0}) and (\ref{inequalityforp0fromgamma}). The upper bounds of $ p_0 $ are calculated from (\ref{In3}),(\ref{In5}),(\ref{Inpre7}), (\ref{dppdrhoatbdry}), (\ref{Inforp0}) \& (\ref{p0otherinequality}). They are listed in Table~\ref{tab:2}. The required lower bound for $ p_0 $ can be as taken as the largest values listed for each $ K $ and the upper bound can be taken as the least values listed for each $ K $. 

\begin{table}[hbtp]
\caption{The upper limits of $ \alpha $ for different permissible values of $ K $.}
\label{tab:1}
\begin{tabular}{cccccccc}
% \toprule
\multicolumn{1}{c}{}&\multicolumn{6}{c}{Inequality Numbers} \\ \cline{2-6}
$ K $ & (\ref{In1}) & (\ref{In3a}) & (\ref{In4a}) & (\ref{In6}) & (\ref{Inpreprepre7}) &   \\ \hline
1.8001 & 0.4898 & 0.0150 & 0.9999 & 7.9882 & 0.1600 \\ 
1.9001 & 0.5244 & 0.0179 & 0.9405 & 7.8029 & 0.2025 \\ 
2.0001 & 0.5556 & 0.0209 & 0.8888 & 7.6282 & 0.2501 \\
2.1001 & 0.5838 & 0.0239 & 0.8439 & 7.4632 & 0.3026 \\ 
2.2001 & 0.6094 & 0.0270 & 0.8047 & 7.3072 & 0.3601 \\ 
2.3001 & 0.6327 & 0.0301 & 0.7704 & 7.1594 & 0.4226 \\ 
2.4001 & 0.6540 & 0.0333 & 0.7405 & 7.0192 & 0.4901 \\ 
2.5001 & 0.6735 & 0.0364 & 0.7143 & 6.8861 & 0.5626 \\ 
2.6001 & 0.6914 & 0.0396 & 0.6913 & 6.7594 & 0.6401 \\
2.7001 & 0.7078 & 0.0427 & 0.6713 & 6.6388 & 0.7226 \\ 
2.8001 & 0.7230 & 0.0459 & 0.6537 & 6.5239 & 0.8101 \\
2.9001 & 0.7370 & 0.0490 & 0.6384 & 6.4142 & 0.9026 \\ 
3.0001 & 0.7500 & 0.0521 & 0.6250 & 6.3093 & 1.0001 \\ 
3.1001 & 0.7621 & 0.0552 & 0.6133 & 6.2091 & 1.1026 \\ 
3.2001 & 0.7733 & 0.0583 & 0.6032 & 6.1131 & 1.2101 \\ 
3.3001 & 0.7837 & 0.0613 & 0.5944 & 6.0211 & 1.3226 \\ 
3.3401 & 0.7876 & 0.0625 & 0.5912 & 5.9853 & 1.3690 \\ \hline
\end{tabular}
\end{table}

Similarly the lower bound for $ p_0 $ can be easily seen from the below Table~\ref{tab:2}.

\begin{table}[hbtp]
\caption{For a fixed value $ 0.05 $ of $ \alpha $ from the above Table \ref{tab:1}, the lower and upper limits of $ p_0 $ for different permissible values of $ K $.}
\label{tab:2}
\begin{tabular}{cccccccccc}
\multicolumn{1}{c}{}&\multicolumn{8}{c}{Inequality Numbers} \\ \cline{2-9}
\multicolumn{1}{c}{}&\multicolumn{2}{c}{Lower Bounds} & \multicolumn{6}{c}{Upper Bounds} \\ \cline{2-3} \cline{{4~~~}-9}
$ K $ & (\ref{Inforp0}) & (\ref{inequalityforp0fromgamma}) & (\ref{In3}) & (\ref{In5}) & (\ref{Inpre7}) & (\ref{dppdrhoatbdry}) & (\ref{Inforp0}) & (\ref{p0otherinequality}) &  \\ \hline
1.8001 & 0.0517 & 0.2852 & -0.4458 & 0.8001 & 2.0280 & 2.1409 & 1.2451 & 1.8315 \\ 
1.9001 & 0.0685 & 0.3176 & -0.4425 & 0.9001 & 2.2766 & 2.4551 & 1.4281 & 2.2249 \\ 
2.0001 & 0.0860 & 0.3500 & -0.4333 & 1.0001 & 2.5252 & 2.7837 & 1.6147 & 2.6409 \\ 
2.1001 & 0.1042 & 0.3826 & -0.4178 & 1.1001 & 2.7741 & 3.1262 & 1.8045 & 3.0756 \\ 
2.2001 & 0.1229 & 0.4152 & -0.3957 & 1.2001 & 3.0230 & 3.4824 & 1.9972 & 3.5253 \\ 
2.3001 & 0.1421 & 0.4479 & -0.3666 & 1.3001 & 3.2720 & 3.8520 & 2.1925 & 3.9860 \\ 
2.4001 & 0.1618 & 0.4806 & -0.3302 & 1.4001 & 3.5211 & 4.2349 & 2.3902 & 4.4537 \\ 
2.5001 & 0.1819 & 0.5134 & -0.2862 & 1.5001 & 3.7702 & 4.6308 & 2.5901 & 4.9247 \\ 
2.6001 & 0.2023 & 0.5462 & -0.2343 & 1.6001 & 4.0195 & 5.0395 & 2.7921 & 5.3952 \\ 
2.7001 & 0.2231 & 0.5790 & -0.1740 & 1.7001 & 4.2688 & 5.4610 & 2.9959 & 5.8619 \\ 
2.8001 & 0.2442 & 0.6119 & -0.1051 & 1.8001 & 4.5181 & 5.8951 & 3.2015 & 6.3216 \\ 
2.9001 & 0.2655 & 0.6449 & -0.0271 & 1.9001 & 4.7675 & 6.3416 & 3.4086 & 6.7715 \\ 
3.0001 & 0.2871 & 0.6778 & 0.0601 & 2.0001 & 5.0169 & 6.8005 & 3.6173 & 7.2091 \\ 
3.1001 & 0.3089 & 0.7108 & 0.1570 & 2.1001 & 5.2664 & 7.2716 & 3.8273 & 7.6325 \\ 
3.2001 & 0.3309 & 0.7438 & 0.2639 & 2.2001 & 5.5159 & 7.7549 & 4.0386 & 8.0398 \\ 
3.3001 & 0.3531 & 0.7768 & 0.3811 & 2.3001 & 5.7654 & 8.2502 & 4.2511 & 8.4296 \\ 
3.3401 & 0.3620 & 0.7900 & 0.4310 & 2.3401 & 5.8652 & 8.4517 & 4.3365 & 8.5804 \\ \hline
\end{tabular}
\end{table}

\section{Application to Compact Stars}
\label{sec:4}

We shall use the charged anisotropic model on pseudo-spheroidal spacetime to strange star models whose mass and size are known. Consider the pulsar 4U 1820-30 whose estimated mass and radius are $ 1.58 M_{\odot} $ and $ 9.1 km $. If we use these estimates in (\ref{M}) with $ \alpha = 0.05 $, we get $ K = 2.718 $  which is well inside the permitted limits of $ K $. Similarly by taking the estimated masses and radii of some well-known pulsars like  PSR J1903+327, 4U 1608-52, Vela X-1, SMC X-4 and Cen X-3, we have calculated the values of $ K $ with $ \alpha = 0.05 $ for each of these stars. These estimates together with some relevant physical quantities like the central density $ \rho_c, $ surface density $ \rho_R, $ the compactification factor $ u = \frac{M}{R}, \frac{dp_r}{d\rho}_{(r = 0)} $ and the charge $ Q $ inside the star are displayed in Table \textbf{\ref{tab:3}}. From this table it is evident that charged anisotropic models can accommodate the observational data of pulsars recently given by \cite{Gangopadhyay13}. \\

\begin{table}[h]
\footnotesize
\caption{Estimated physical values based on the observational data with $ \alpha = 0.05 $ fixed.}
\label{tab:3}
\begin{tabular}{lllllllll}
\hline\noalign{\smallskip}
\textbf{STAR} & $\mathbf{K} $ & {$ \mathbf{M} $} & {$ \mathbf{R} $} & {$ \mathbf{\rho_c} $} & {$ \mathbf{\rho_R} $} & {$ \mathbf{u (=\frac{M}{R})} $} & $ \mathbf{\left(\frac{dp_r}{d \rho}\right)_{r=0}} $ & {$\mathbf{Q}$} \\
& & $ \mathbf{(M_\odot)} $ & $ \mathbf{(Km)} $ & \textbf{(MeV fm{$\mathbf{^{-3}}$})} & \textbf{(MeV fm{$\mathbf{^{-3}}$})} & & & \textbf{Coulombs} \\
\noalign{\smallskip}\hline\noalign{\smallskip}
\textbf{4U 1820-30} 	  & 2.718 & 1.58  & 9.1   & 1875.15 & 240.29 & 0.256 & 0.251 & 2.36 $ \times 10^{20} $ \\
\textbf{PSR J1903+327} 	  & 2.781 & 1.667 & 9.438 & 1806.35 & 226.58 & 0.261 & 0.242 & 2.45 $ \times 10^{20} $ \\
\textbf{4U 1608-52} 	  & 3.010 & 1.74  & 9.31  & 2095.78 & 243.66 & 0.276 & 0.214 & 2.42 $ \times 10^{20} $ \\
\textbf{Vela X-1} 	  & 2.969 & 1.77  & 9.56  & 1947.00 & 229.38 & 0.273 & 0.218 & 2.48 $ \times 10^{20} $ \\
\textbf{SMC X-4}          & 2.230 & 1.29  & 8.831 & 1425.58 & 218.84 & 0.300 & 0.350 & 2.29 $ \times 10^{20} $ \\
\textbf{Cen X-3}          & 2.502 & 1.49  & 9.178 & 1610.99 & 223.03 & 0.239 & 0.287 & 2.38 $ \times 10^{20} $ \\
\noalign{\smallskip}\hline
\end{tabular} 
\end{table}
\section{Validation of Model for 4U 1820-30}
\label{sec:5}
\begin{figure}
\includegraphics[scale = 0.5]{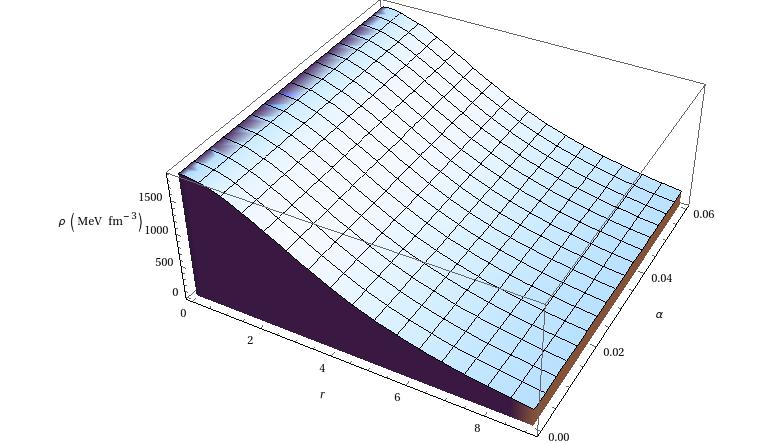} 
\caption{Variation of density against radial parameter $r$ and charge parameter $ \alpha $ for $ K = 2.718 $. 
\label{fig:1}}
\end{figure}

\begin{figure}
\includegraphics[scale = 0.5]{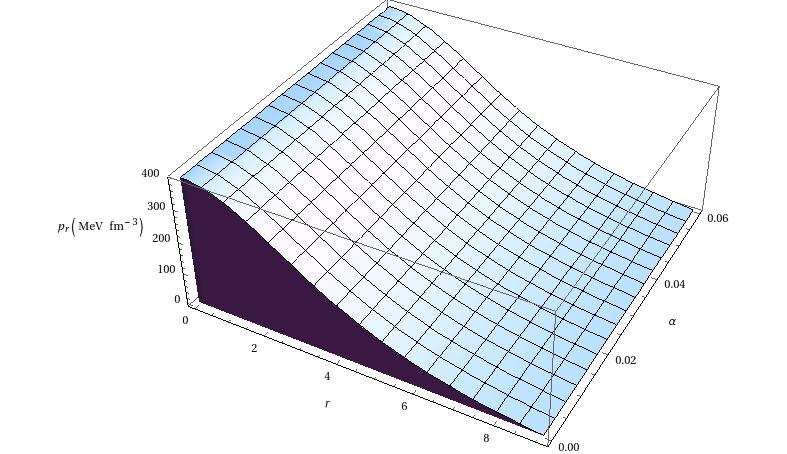}
\caption{Variation of radial pressures against radial parameter $r$ and charge parameter $ \alpha $ for $ K = 2.718 $.
\label{fig:2}}
\end{figure}

\begin{figure}
\includegraphics[scale = 0.5]{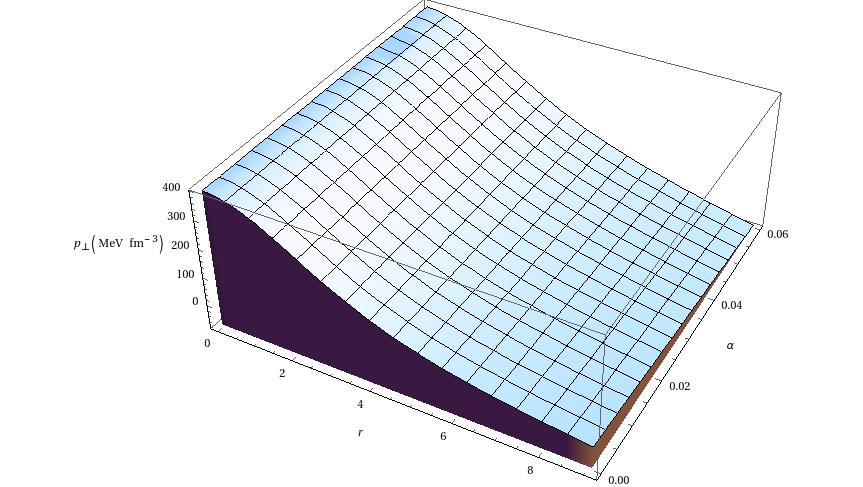}
\caption{Variation of transverse pressures against radial parameter $r$ and charge parameter $ \alpha $ for $ K = 2.718 $. 
\label{fig:3}}
\end{figure}

In order to examine the nature of physical quantities throughout the distribution, we have considered a particular pulsar 4U 1820-30, whose tabulated mass and radius are $M=1.58 M_{\odot},$ and $ R=9.1 km $ respectively. From Table~\ref{tab:3} it can be noticed that the corresponding values of $ K=2.718 $ with $ \alpha = 0.05 $. We have shown the variations of density and pressures in both the charged and uncharged cases in Figure~\ref{fig:1}, Figure~\ref{fig:2} and Figure~\ref{fig:3}. It can be noticed that the density is decreasing radially outward. Similarly the radial pressure $p_{r}$ and transverse pressure $p_{\perp}$ are decreasing radially outward.

\begin{figure}
\includegraphics[scale = 0.5]{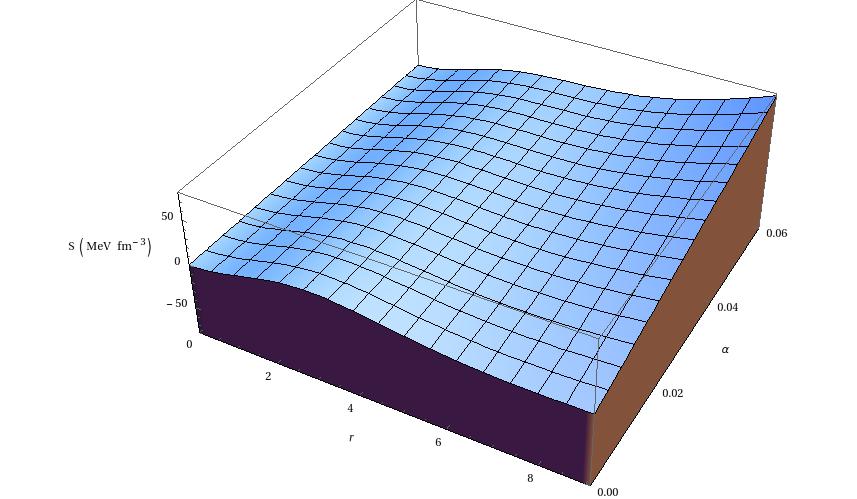}
\caption{Variation of anisotropy against radial parameter $r$ and charge parameter $ \alpha $ for $ K = 2.718 $. 
\label{fig:4}}
\end{figure}

\begin{figure}
\includegraphics[scale = 0.5]{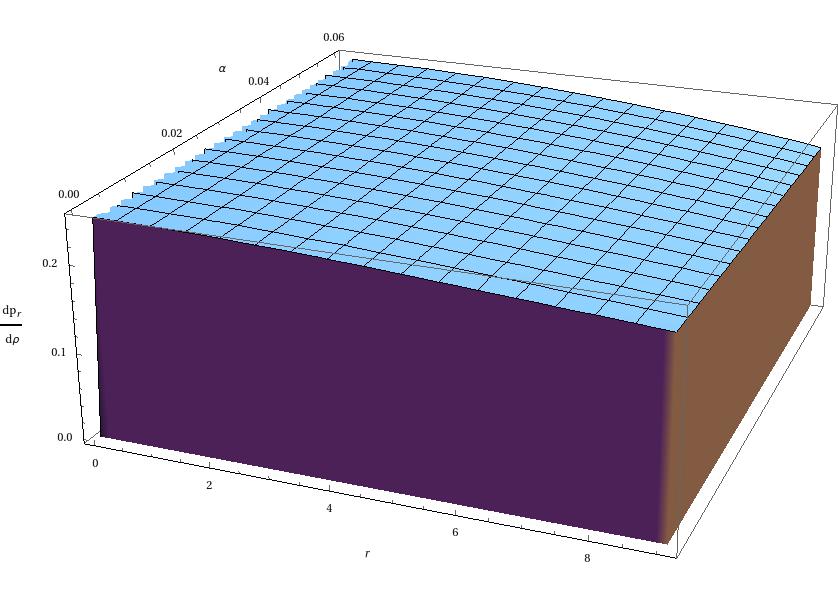}
\caption{Variation of $ \frac{1}{c^2}\frac{dp_r}{d\rho} $ against radial parameter $r$ and charge parameter $ \alpha $ for $ K = 2.718 $. 
\label{fig:5}}
\end{figure}

\begin{figure}
\includegraphics[scale = 0.5]{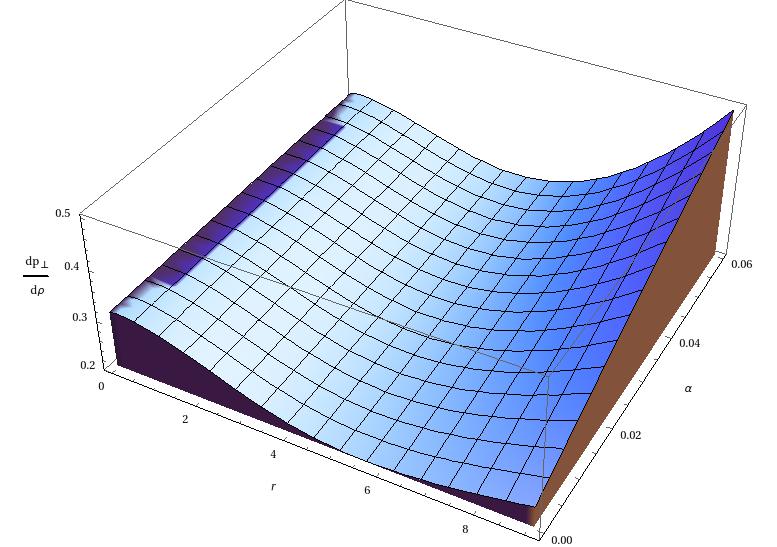}
\caption{Variation of $ \frac{1}{c^2}\frac{dp_\perp}{d\rho} $ against radial parameter $r$ and charge parameter $ \alpha $ for $ K = 2.718 $.
\label{fig:6}}
\end{figure}

\begin{figure}
\includegraphics[scale = 0.51]{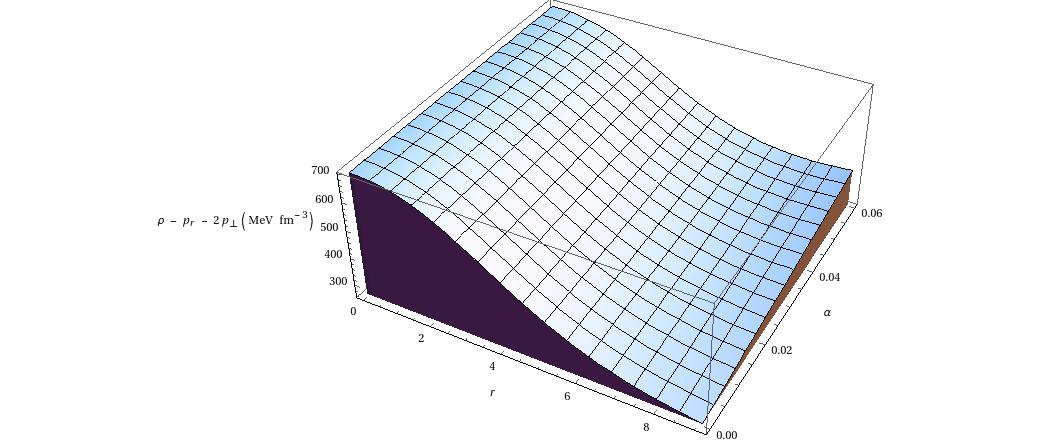}
\caption{Variation of strong energy condition against radial parameter $r$ and charge parameter $ \alpha $ for $ K = 2.718 $. 
\label{fig:7}}
\end{figure}

The variation of anisotropy shown in Figure~\ref{fig:4} is initially decreasing with negative values reaches minimum and then increases. The square of sound speed in the radial and transverse direction (i.e. $\frac{dp_{r}}{d\rho}$ and $\frac{dp_{\perp}}{d\rho}$) are shown in Figure~\ref{fig:5} and 
Figure~\ref{fig:6} respectively and found that they are less than 1, showing that the causality condition is fulfilled throughout. The graph of $\rho-p_{r}-2p_{\perp}$ against radius is plotted in Figure~\ref{fig:7}. It can be observed that it is non-negative for $ 0 \leq r \leq R $ and hence strong energy condition is satisfied throughout the star. \\

\begin{figure}
\includegraphics[scale = 0.5]{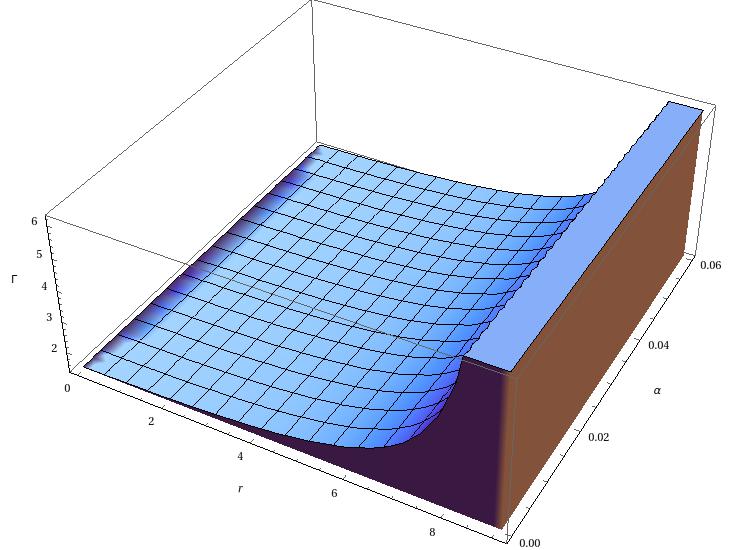}
\caption{Variation of $ \Gamma $ against radial parameter $r$ and charge parameter $ \alpha $ for $ K = 2.718 $. 
\label{fig:8}}
\end{figure}

\begin{figure}
\includegraphics[scale = 0.5]{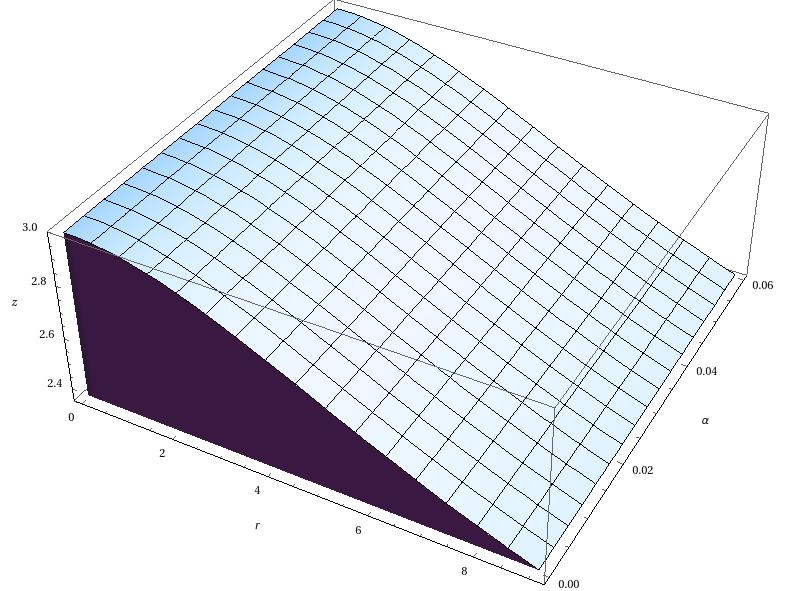}
\caption{Variation of gravitational redshift against radial parameter $r$ and charge parameter $ \alpha $ for $ K = 2.718 $. 
\label{fig:9}}
\end{figure}

\begin{figure}
\includegraphics[scale = 0.5]{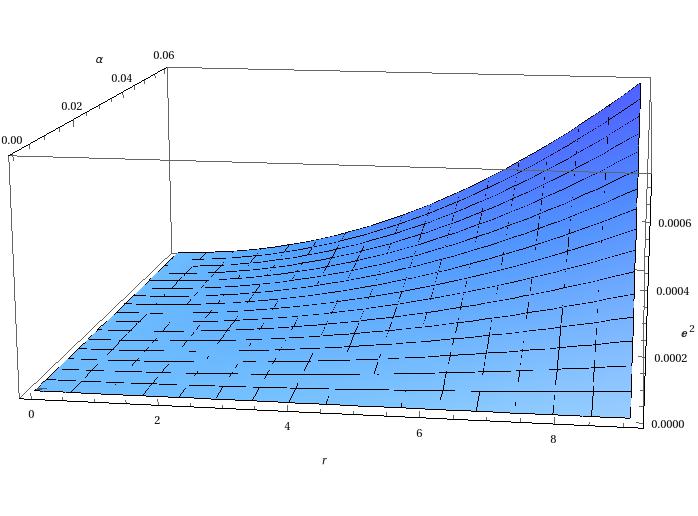}
\caption{Variation of $ E^2 $ against radial parameter $r$ and charge parameter $ \alpha $ for $ K = 2.718 $. 
\label{fig:10}}
\end{figure}

A necessary condition for the exact solution to represent stable relativistic star is that the relativistic adiabatic index given by $ \Gamma = \frac{\rho + p_r}{p_r} \frac{d p_r}{d \rho} $ should be greater than $ \frac{4}{3}. $ The variation of adiabatic index throughout the star is shown in Figure~\ref{fig:8} and it is found that $ \Gamma > \frac{4}{3} $ (\cite{Knutsen87}) throughout the distribution both in charged and uncharged case. For a physically acceptable relativistic star the gravitational redshift must be positive and finite at the centre and on the boundary. Further it should be a decreasing function of $ r $ (\cite{Murad13}). Figure~\ref{fig:9} shows that this is indeed the case. For a physically acceptable charged distribution, the electric field intensity $ E $ should be an increasing function of $ r $ (\cite{Murad13}). The variation of $ E^2 $ against $ r $ is displayed in Figure~\ref{fig:10}. $ E^2 $ is found to be radially increasing throughout the distribution. The model reduces to the uncharged anisotropic distribution given by \cite{Thomas15} when $ \alpha = 0. $ \\
\section*{Acknowledgement}
The authors would like to thank IUCAA, Pune for the facilities and hospitality provided to them for carrying out this work.

\end{document}